\documentclass{article}
\usepackage{spconf,amsmath,graphicx,hyperref}
\usepackage{amsfonts}
\usepackage[dvipsnames]{xcolor}
\usepackage{multirow}
\usepackage{rotating}

\usepackage{amsmath,amsfonts,amssymb,bm}

\newcommand{\mbB}{\mathbf{B}}

\newcommand{\mbX}{\mathbf{X}}

\newcommand{\mbx}{\mathbf{x}}
\newcommand{\mby}{\mathbf{y}}
\newcommand{\mbz}{\mathbf{z}}

\newcommand{\mbep}{\boldsymbol{\epsilon}}

\title{GLA-Grad++: An Improved Griffin-Lim Guided Diffusion Model \\for Speech Synthesis}

\name{Teysir Baoueb\textsuperscript{1}, Xiaoyu Bie\textsuperscript{1}, Mathieu Fontaine\textsuperscript{1}, Gaël Richard\textsuperscript{1} \thanks{With support from the European Union (ERC, HI-Audio - Hybrid and Interpretable Deep neural
audio machines, 101052978).
Views and opinions expressed are however those of the author(s) only and do not necessarily reflect those
of the European Union or the European Research Council. Neither the European Union nor the granting
authority can be held responsible for them.}}
\address{$^{1}$LTCI, T\'el\'ecom Paris, Institut polytechnique de Paris, France}
\begin{document}
\ninept
\maketitle
\begin{abstract} 
Recent advances in diffusion models have positioned them as powerful generative frameworks for speech synthesis, demonstrating substantial improvements in audio quality and stability. Nevertheless, their effectiveness in vocoders conditioned on mel spectrograms remains constrained, particularly when the conditioning diverges from the training distribution. 
The recently proposed GLA-Grad model introduced a phase-aware extension to the WaveGrad vocoder that integrated the Griffin-Lim algorithm (GLA) into the reverse process to reduce inconsistencies between generated signals and conditioning mel spectrogram. 
In this paper, we further improve GLA-Grad through an innovative choice in how to apply the correction. Particularly, we compute the correction term only once, with a single application of GLA, to accelerate the generation process. Experimental results demonstrate that our method consistently outperforms the baseline models, particularly in out-of-domain scenarios.
\end{abstract}
\begin{keywords}
Diffusion models, speech generation, speech vocoder, Griffin-Lim algorithm, domain adaptation
\end{keywords}
\section{Introduction}
\label{sec:intro}
Generative modeling~\cite{goodfellow2014generative, rezende2015variational, ho2020denoising} has seen remarkable advances in recent years, enabling the synthesis of high-quality and diverse data across domains such as vision and speech~\cite{dhariwal2021diffusion, chen2020wavegrad}. 
Among the emerging approaches, diffusion models~\cite{ho2020denoising, song2021scorebased} have attracted considerable attention due to their stability, flexibility, and ability to produce realistic samples.  
While originally developed for image synthesis, these models have quickly been adapted to audio tasks, including speech generation~\cite{scheibler2023diffusion}, enhancement~\cite{lemercier2023storm, richter2023speech}, and separation~\cite{chen2023sepdiff}. 
Diffusion-based vocoders, such as WaveGrad~\cite{chen2020wavegrad} and DiffWave~\cite{kong2020diffwave}, as well as their extensions PriorGrad~\cite{lee2021priorgrad} and SpecGrad~\cite{koizumi2022specgrad}, demonstrate the potential of this framework to produce high-fidelity audio conditioned on mel spectrograms.    

Despite these successes, diffusion-based vocoders still face two key challenges. 
First, the iterative reverse process is computationally expensive, which limits real-time applications and practical deployment~\cite{huang2022fastdiff}. 
Second, robustness to unseen conditions is limited: mismatches in the conditioning mel spectrogram or in speaker characteristics often result in degraded audio quality~\cite{farahani2020brief}. 
In fact, accurate reconstruction of both phase and magnitude is crucial for natural-sounding speech, but conventional methods may produce inconsistencies, particularly in out-of-domain scenarios. 

To address these challenges, several recent vocoders have explored alternative strategies. 
FreGrad~\cite{nguyen2024fregradlightweightfastfrequencyaware} introduces a frequency-aware wavelet decomposition and lightweight architecture to achieve high-fidelity speech synthesis with significantly reduced training and inference costs. 
DiffPhase~\cite{peer2023diffphase} trains a diffusion model specifically for STFT phase retrieval. 
PeriodGrad~\cite{hono2024periodgradpitchcontrollableneuralvocoder}, PeriodWave~\cite{lee2025periodwave}, and PeriodWave-Turbo~\cite{lee2024acceleratinghighfidelitywaveformgeneration} explicitly capture periodic structures using embeddings or multi-period flow matching, improving pitch and waveform fidelity. 
Cauchy Diffusion~\cite{lian2025cauchy} addresses mode collapse arising from imbalanced speech data by adopting heavy-tailed distributions. 
Finally, SWave~\cite{liu2025swave} introduces rectified flow with a three-stage training process to reduce the number of generation steps while maintaining high audio quality.  

A common feature of these approaches is that they generally require retraining to achieve their improvements. 
In contrast to GLA-Grad,~\cite{liu2024glagradgriffinlimextendedwaveform} proposed a phase-aware extension of WaveGrad that integrates the Griffin-Lim algorithm (GLA)~\cite{griffin1984signal} into the reverse diffusion process in a zero-shot manner.
This enables immediate application to unseen conditioning signals while reducing inconsistencies between the generated waveform and the mel spectrogram. 
This approach improved robustness to mismatched inputs but still requires multiple iterations of GLA for the first steps of the diffusion process. 

\begin{figure}[t]
\centering
\includegraphics[width=0.98\columnwidth]{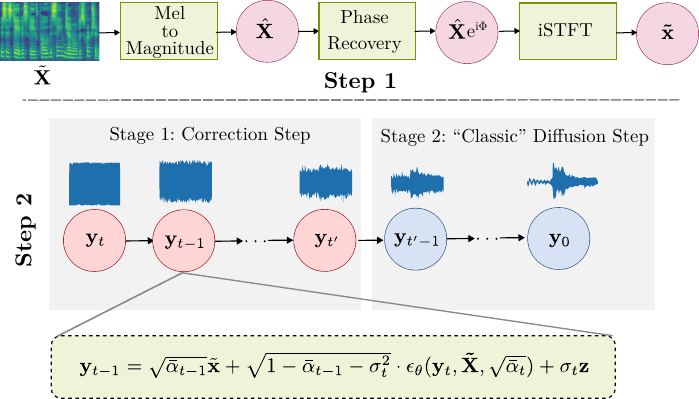}
\caption{Overview of GLA-Grad++: Step 1 (top): Before starting the diffusion process, we estimate the audio from the mel spectrogram; Step 2 (bottom): We run the reverse diffusion process, where we use $\tilde{\mathbf{x}}$ in Stage 1 to correct the predicted $\mathbf{y}_0$, and then switch to the classical diffusion process in Stage 2.}
\vspace{-.3cm}
\label{fig:method}
\end{figure}

In this paper, we enhance GLA-Grad by computing the correction term only once during generation and applying it at a carefully chosen point in the reverse diffusion process. 
This approach, which we call GLA-Grad++, simultaneously improves robustness to out-of-domain inputs and reduces computational overhead, accelerating the generation process. Audio examples are available on our demo page at the following link: \href{https://gla-grad-plus-plus.github.io/}{https://gla-grad-plus-plus.github.io/}.

\section{Background}\label{sec:background}
\subsection{Diffusion models}
Denoising Diffusion Probabilistic models (DDPM)~\cite{ho2020denoising} are a class of generative models that learn to transform simple noise distributions into complex data distributions through a sequence of gradual denoising steps. DDPMs are composed of two processes: the forward process and the reverse process. During the former, the data is progressively corrupted with Gaussian noise. Formally, the forward process is defined as:

\begin{equation}
q(\mby_t|\mby_{t-1}) = \mathcal{N}\big(\mby_t; \sqrt{1-\beta_t} \, \mby_{t-1}, \, \beta_t \mathbf{I}\big),
\end{equation}
where $\beta_t$ controls the noise level at step $t$, and this process eventually transforms the datum $\mby_0$ into nearly isotropic Gaussian noise $\mby_T$. 
Instead of applying the previous equation iteratively from $\mby_0$, the equation could be further simplified to:

\begin{equation}
    \mby_t = \sqrt{\bar{\alpha}_t} \mby_0 + \sqrt{1-\bar{\alpha}_t} \mbep, \quad \mbep \sim \mathcal{N}(\mathbf{0}, \mathbf{I}), \label{eq:forward_process}
\end{equation}
where $\alpha_t = 1-\beta_t$, $\bar{\alpha}_t = \prod_{s=1}^t \alpha_s$.

The generative task is then to learn a reverse process that iteratively denoises $\mby_t$ to recover $\mby_0$. In the DDPM framework, the reverse process can be written as:

\begin{equation}
\resizebox{0.91\hsize}{!}{
$\mby_{t-1} = \frac{1}{\sqrt{\alpha_t}} \Big( \mby_t - \frac{1-\alpha_t}{\sqrt{1-\bar{\alpha}_t}} \, \mbep_\theta(\mby_t,t) \Big) + \sigma_t \mbz, \quad \mbz \sim \mathcal{N}(\mathbf{0},\mathbf{I}),$}
\end{equation}
where $\mbep_\theta(\mby_t,t)$ is a neural network that predicts the noise in $\mby_t$, and $\sigma_t = \sqrt{\frac{1 - \bar{\alpha}_{t - 1}}{1 - \bar{\alpha}_t}\beta_t}$ controls the variance of the added Gaussian noise. The training objective is a simplified mean-squared error loss on the predicted noise:

\begin{equation}
\mathcal{L}_{\text{DDPM}} = \mathbb{E}_{\mby_0,\mbep \sim \mathcal{N}(\mathbf{0},\mathbf{I}),t} \big[ \| \mbep - \mbep_\theta(\mby_t, t) \|_2^2 \big].
\end{equation}

WaveGrad~\cite{chen2020wavegrad} is a DDPM-based vocoder that generates audio waveform conditioned on mel spectrogram. Compared to the standard DDPM framework, it uses the $L_1$ norm in the loss and uses the following denoising equation:
\begin{equation}
\mby_{t-1}=\frac{1}{\sqrt{\alpha_t}} \left(\mby_t-\frac{1-\alpha_t}{\sqrt{1-\bar{\alpha}_t}} \mbep_\theta\left(\mby_t, \tilde{\mbX}, \sqrt{\bar{\alpha}_t}\right)\right) + \sigma_t \mbz, \label{eq:wavegrad_gen}
\end{equation}
where $\tilde{\mathbf{X}}$ is the conditioning mel spectrogram. WaveGrad~\cite{chen2020wavegrad} reparameterized the model to condition on a continuous noise level.

\subsection{Griffin-Lim algorithm}

The Griffin-Lim algorithm~\cite{griffin1984signal} is a widely used iterative method for reconstructing a time-domain signal from a magnitude spectrogram when the phase information is unavailable. The key idea is to estimate a signal whose Short-Time Fourier Transform (STFT) magnitude matches the given spectrogram while iteratively refining the phase. Let $|X(\omega, t)|$ denote the target magnitude spectrogram and $\mathcal{G}\{x[n]\}$ denote the STFT of a signal $x[n]$. The algorithm alternates between the following two projection steps:

\begin{enumerate}
    \item \textbf{Projection onto the magnitude constraint:} Given a complex STFT estimate $Y_k(\omega, t)$ at iteration $k$, the updated STFT is obtained by replacing its magnitude with the target magnitude while preserving the current phase:
    \begin{equation}
    Y_k'(\omega, t) = |X(\omega, t)| \frac{Y_k(\omega, t)}{|Y_k(\omega, t)|}.
    \end{equation}

    \item \textbf{Projection onto the time-domain consistency:} Convert the modified STFT back to the time domain using the inverse STFT, then recompute the STFT of the resulting signal to enforce consistency between overlapping frames:
    \begin{equation}
    Y_{k+1}(\omega, t) = \mathcal{G}\big\{ \mathcal{G}^{-1}\{ Y_k'(\omega, t) \} \big\}.
    \end{equation}
\end{enumerate}

The algorithm is initialized with a random phase and iterated until convergence, producing a time-domain signal $x[n]$ whose STFT magnitude approximates the target $|X(\omega, t)|$. Griffin-Lim can also be interpreted as a projection onto the intersection of two constraint sets: one defined by the target magnitude and the other by the consistency of the STFT synthesis, guaranteeing convergence to a local minimum in the Euclidean distance between the magnitude spectrograms.

\section{Proposed Method}\label{sec:method}

In the reverse diffusion process, we begin with random Gaussian noise and iteratively denoise it to obtain the generated waveform. During the early iterations, the intermediate samples $\mby_t$ remain highly noisy. Following~\cite{song2022denoisingdiffusionimplicitmodels}, Equation~\ref{eq:wavegrad_gen} can be reformulated as:

\begin{align}
    \mby_{t-1} & = \sqrt{\bar{\alpha}_{t-1}} \underbrace{\left(\frac{\mby_t - \sqrt{1 - \bar{\alpha}_t} \mbep_\theta\left(\mby_t, \tilde{\mbX}, \sqrt{\bar{\alpha}_t}\right)}{\sqrt{\bar{\alpha}_t}}\right)}_{\text{``predicted } \mby_0 \text{''}} \notag \\
    & + \underbrace{\sqrt{1 - \bar{\alpha}_{t-1} - \sigma_t^2} \cdot \mbep_\theta\left(\mby_t, \tilde{\mbX}, \sqrt{\bar{\alpha}_t}\right)}_{\text{``direction pointing to } \mby_t \text{''}} + \underbrace{\sigma_t \mbz}_{\text{random noise}} \label{eq:ddim_gen}
\end{align}

Each term in the previous update admits a clear interpretation. The first term rescales the denoised estimate (or the predicted $\mby_0$), an interpretation that follows directly from rearranging the terms in Equation~\ref{eq:forward_process}. The second term accounts for the contribution of the predicted noise at the current step. The third reintroduces controlled stochasticity. 

Motivated by this observation, we hypothesize that replacing the predicted $\mby_0$ with a more accurate estimate during the early iterations of the denoising process could improve the overall efficiency of generation. 

\subsection{Magnitude spectrogram estimation}\label{subsec:magnitude}
The model is conditioned on the mel spectrogram $\tilde{\mbX}$, which serves as a lossy representation of the magnitude spectrogram $\mbX$. It is obtained using the following transformation: $\tilde{\mbX} = \mbB \mbX$, where $\mbB \in \mathbb{R}_+^{M \times F}$ denotes the mel filterbank matrix, $F$ is the number of frequency bins, and $M < F$ is the number of mel bands.

To reconstruct the magnitude spectrogram from the mel spectrogram, we apply the pseudo-inverse of the mel filterbank to $\tilde{\mbX}$: $\hat{\mbX} = \mbB^{+} \tilde{\mbX}$.

\subsection{Phase recovery}
Phase estimation for the magnitude spectrogram is performed using the Griffin-Lim algorithm (GLA). Unlike~\cite{liu2024glagradgriffinlimextendedwaveform}, we do not initialize the phase with the current iterate; instead, we use random initialization. This choice is motivated by the observation that, during the initial iterations, the phase information is unreliable and has little influence on the convergence of GLA. Consequently, in our approach, GLA is independent of the diffusion process and does not need to be applied at every diffusion step; it can be applied just once prior to the denoising process.

After recovering the magnitude and phase, we use the inverse STFT to reconstruct the time-domain signal $\tilde{\mbx}$. Similar to~\cite{liu2024glagradgriffinlimextendedwaveform}, our generation process consists of two stages. However, in the first stage, we do not update the entire $\mby_t$. Instead, we replace the predicted $\mby_0$ term in Equation~\ref{eq:ddim_gen} with $\tilde{\mbx}$ during the initial steps of the diffusion process. The new equation is given by:
\begin{align}
    \resizebox{0.91\hsize}{!}{$\mby_{t-1} = \sqrt{\bar{\alpha}_{t-1}} \tilde{\mbx}
    + \sqrt{1 - \bar{\alpha}_{t-1} - \sigma_t^2} \cdot \mbep_\theta\left(\mby_t, \tilde{\mbX}, \sqrt{\bar{\alpha}_t}\right)+ \sigma_t \mbz$}
    \label{eq:eq_gen}
\end{align}

In the second stage, as the denoised estimate improves over later diffusion steps, we revert to the standard denoising equation (Equation~\ref{eq:ddim_gen}). This process is illustrated in Figure~\ref{fig:method}.

\section{Experiments}\label{sec:experiments}
\subsection{Datasets}
For our experiments, we make use of the following datasets:

\begin{itemize}
\item \textbf{LJSpeech}~\cite{ljspeech17} is a single-speaker English speech dataset consisting of recordings sampled at $22050$ Hz, with a total duration of approximately $24$ hours. Following the protocol of HiFi-GAN~\cite{kong2020hifigan}, we adopt the same data split: $12{,}950$ clips for training and $150$ clips for testing.
\item \textbf{VCTK v0.92}~\cite{Yamagishi2019CSTRVC} is a clean, multi-speaker dataset featuring $110$ speakers ($63$ female, $47$ male). This version improves upon the original VCTK by adding one additional speaker and removing silent segments and faulty tracks. Recordings were made with two microphones, and we used the Microphone $1$ setup. The corpus comprises roughly $41$ hours of English utterances across diverse accents. We downsample the original $48$ kHz recordings to $24$ kHz. For evaluation, we retain the same $10$ speakers as~\cite{baoueb2024specdiffganspectrallyshapednoisediffusion} to ensure a balanced and diverse test set in terms of accent, gender, and age, while the remaining speakers are used for training.
\end{itemize}

\subsection{Metrics}
The evaluation relies on three metrics: 
\begin{itemize}
    \item \textbf{Perceptual Evaluation of Speech Quality (PESQ)}~\cite{pesq}, 
    a standardized method designed for automated assessment of speech quality as perceived by users in telephony applications. 
    \item \textbf{Short Term Objective Intelligibility (STOI)}~\cite{stoi}, 
    a measure intended to estimate the intelligibility of speech in noisy conditions. 
    \item \textbf{WARP-Q}~\cite{warpQ}, 
    an objective speech quality metric that employs a subsequence dynamic time warping (SDTW) algorithm on MFCC features to produce a raw quality score reflecting the similarity between the generated and the reference speech signals. 
\end{itemize}

\subsection{Model setup}
For our experiments, we evaluate the following baselines: WaveGrad with WG-6 and GLA-Grad. Here, WG-6 denotes optimally selected noise schedules for WaveGrad with 6 iterations.

For all methods, we use the following parameters for mel spectrogram computation: $n_{\text{fft}} = 2048$, a Hann window of length $1200$, a hop size of $300$, and $n_{\text{mels}} = 128$.  For both GLA-Grad and GLA-Grad++, we use the Fast GLA implementation with $32$ iterations. We train the WaveGrad models on LJSpeech and VCTK for 1M steps.

For the oracle setup, we first use the ground-truth magnitude spectrogram with Griffin-Lim algorithm to evaluate how an improved spectrogram affects the generation. Next, we combine the ground-truth phase with the reconstructed magnitude spectrogram to assess the impact of phase on the generation.

Unless otherwise specified, we use the DDPM $\sigma$ from Equation~\ref{eq:ddim_gen} with the WG-6 noise schedule. Following~\cite{liu2024glagradgriffinlimextendedwaveform}, each stage consists of three steps. 
As will be shown in Sections~\ref{sec:impact_end_timestep} and~\ref{sec:impact_end_timestep_per_file}, this choice proves suboptimal for our method.

\section{Results}\label{sec:results}
\subsection{Oracle results}
Table~\ref{tab:results_magnitude_oracle} reports the oracle results, assuming that each stage in~\ref{fig:method} spans three steps. 
In both scenarios, i.e., when the real magnitude spectrogram is provided or when the real phase is given, we observe a positive impact on the generation process. Notably, having access to the correct phase produces better results than having access to the proper magnitude spectrogram. This confirms  the interest of using a more accurate predicted $\mby_0$ for diffusion inference.

\begin{table}[ht]
    \centering
    \resizebox{\columnwidth}{!}{
    \begin{tabular}{c|c|c|c|c}
         \multirow{4}{*}{\begin{turn}{90}LJSpeech\end{turn}} & \textbf{Approach} & \textbf{PESQ} ($\uparrow$) & \textbf{STOI} ($\uparrow$) & \textbf{WARP-Q} ($\downarrow$) \\
         & WaveGrad & $3.598 \pm 0.127$ & $0.970 \pm 0.005$ & $1.665 \pm 0.078$ \\
         & Oracle Spec & $3.892 \pm 0.113$ & $0.978 \pm 0.004$ & $1.684 \pm 0.074$ \\
         & Oracle Phase & $4.040 \pm 0.103$ & $0.987 \pm 0.003$ & $1.587 \pm 0.082$ \\
         \hline
        \multirow{3}{*}{\begin{turn}{90}VCTK\end{turn}} & WaveGrad & $3.453 \pm 0.325$ & $0.907 \pm 0.055$ & $1.439 \pm 0.100$\\
         & Oracle Spec & $3.866 \pm 0.219$ & $0.921 \pm 0.056$ & $1.433 \pm 0.099$ \\
         & Oracle Phase & $4.041 \pm 0.238$ & $0.927 \pm 0.057$ & $1.340 \pm 0.104$ \\
    \end{tabular}
    }
    \caption{Results using the ground-truth spectrogram/phase (mean with standard deviation)}
    \label{tab:results_magnitude_oracle}
\end{table}

\subsection{Comparison between different methods}
In Table~\ref{tab:results_different_methods}, we compare the performance of the different methods on LJSpeech and VCTK. The results demonstrate that GLA-Grad++ outperforms the baselines in terms of PESQ and STOI, while achieving comparable scores on WARP-Q.

This overall improvement can be attributed to its more effective use of information from the conditioning mel spectrogram compared to WaveGrad. 
Moreover, unlike GLA-Grad, which replaces the entire iterate $\mby_t$, GLA-Grad++ only replaces the predicted $\mby_0$ term. 
This approach is more principled, as both the predicted $\mby_0$ and the reconstructed audio $\tilde{\mbx}$ have the same interpretation, i.e., correspond to clean speech. 
Additionally, incorporating the required noise level for the current iteration via the directional term pointing to $\mby_t$ ensures consistency in the diffusion process.

The results for GLA-Grad for VCTK are different from those in \cite{liu2024glagradgriffinlimextendedwaveform} due to different training settings.

\begin{table}[ht]
    \centering
    \resizebox{\columnwidth}{!}{
    \begin{tabular}{c|c|c|c|c}
         \multirow{4}{*}{\begin{turn}{90}LJSpeech\end{turn}} & \textbf{Approach} & \textbf{PESQ} ($\uparrow$) & \textbf{STOI} ($\uparrow$) & \textbf{WARP-Q} ($\downarrow$) \\
         & WaveGrad & $3.598 \pm 0.127$ & $0.970 \pm 0.005$ & $\mathbf{1.665} \pm 0.078$\\
         & GLA-Grad & $3.460 \pm 0.112$ & $0.963 \pm 0.005$ & $1.677 \pm 0.076$ \\
         & GLA-Grad++ & $\mathbf{3.807} \pm 0.115$ & $\mathbf{0.974} \pm 0.004$ & $1.694 \pm 0.079$ \\
         \hline
        \multirow{4}{*}{\begin{turn}{90}VCTK\end{turn}} & WaveGrad & $3.453 \pm 0.325$ & $0.907 \pm 0.055$ & $\mathbf{1.439} \pm 0.100$ \\
         & GLA-Grad & $2.024 \pm 0.189$ & $0.858 \pm 0.087$ & $1.758 \pm 0.163$ \\
         & GLA-Grad++ & $\mathbf{3.772} \pm 0.228$ & $\mathbf{0.917} \pm 0.057$ & $1.443 \pm 0.098$ \\
    \end{tabular}
    }
    \caption{Results for the different methods on LJSpeech and VCTK}
    \label{tab:results_different_methods}
\end{table}

We also evaluated the methods using the DDIM value for $\sigma$ (i.e., $\sigma=0$, corresponding to a deterministic process) and the proposed method achieved consistently better results across all three metrics, with a larger margin of improvement. 
Overall, the results obtained with the DDPM $\sigma$ were superior and have therefore been included in this paper.

\subsection{Time complexity results}
We evaluate the inference speed of the different methods compared to real time, using a batch size of $100$ one-second files from the respective test sets of the trained models on a single NVIDIA V100 GPU. The results are presented in~\ref{tab:inference_speed}.

GLA-Grad++ achieves faster inference than GLA-Grad because it applies GLA only once before starting the diffusion process, whereas GLA-Grad applies GLA at each iteration of Stage 1, as it requires the phase of the current iterate $\mby_t$ for initialization. 
Compared to WaveGrad, GLA-Grad++ introduces minimal computational overhead, maintaining efficient and swift inference.

\begin{table}[ht]
    \centering
    \begin{tabular}{c|c|c}
        \textbf{Approach} & \textbf{LJSpeech} & \textbf{VCTK}\\
        WaveGrad & $42.02$ & $39.53$\\
        GLA-Grad & $32.98$ & $31.25$\\
        GLA-Grad++ & $37.80$ & $35.43$ \\
    \end{tabular}
    \caption{Inference Speed compared to real time $(\uparrow)$}
    \label{tab:inference_speed}
\end{table}

\subsection{Impact of the end timestep of stage 1}\label{sec:impact_end_timestep}
For WG-6, we perform $6$ iterations with $t$ decreasing from $6$ to $0$. 
In this section, we explore how varying the endpoint of Stage 1 affects the overall process. 
Notably, the boundary values of this interval (i.e., $0$ and $6$) correspond to the Griffin-Lim algorithm (GLA) (i.e., omitting the second stage), and WaveGrad (i.e., skipping the first stage), respectively. 
The results for LJSpeech and VCTK are reported in Tables~\ref{tab:results_timesteps_ljspeech} and~\ref{tab:results_timesteps_vctk}.

Overall, we observe that ending the first stage at step $2$ yields optimal quality reflected by PESQ scores for both VCTK and LJSpeech. 

In contrast, WARP-Q achieves better results when only GLA is applied. This can likely be attributed to WARP-Q's reliance on MFCCs for computing its score: since GLA modifies only the phase and leaves the magnitude spectrogram unchanged, unlike the diffusion model, which alters both, GLA tends to produce higher WARP-Q scores. 

Regarding STOI, results for LJSpeech remain comparable across the first three rows, while VCTK shows a slight decrease. This drop warrants further investigation to determine its underlying cause.

\begin{table}[ht]
    \centering
    \resizebox{\columnwidth}{!}{
    \begin{tabular}{c|c|c|c}
         \textbf{Approach} & \textbf{PESQ} ($\uparrow$) & \textbf{STOI} ($\uparrow$) & \textbf{WARP-Q} ($\downarrow$) \\
         0 (GLA) & $3.776 \pm 0.125$ & $\mathbf{0.986} \pm 0.002$ & $\mathbf{1.182} \pm 0.050$ \\
         1 & $3.837 \pm 0.112$ & $0.985 \pm 0.002$ & $1.356 \pm 0.065$ \\
         2 & $\mathbf{3.892} \pm 0.098$ & $0.982 \pm 0.003$ & $1.558 \pm 0.074$ \\
         3 & $3.807 \pm 0.115$ & $0.974 \pm 0.004$ & $1.694 \pm 0.079$ \\
         4 & $3.700 \pm 0.125$ & $0.970 \pm 0.004$ & $1.689 \pm 0.076$ \\
         5 & $3.692 \pm 0.117$ & $0.973 \pm 0.004$ & $1.651 \pm 0.075$ \\
         6 (WaveGrad) & $3.598 \pm 0.127$ & $0.970 \pm 0.005$ & $1.665 \pm 0.078$ \\
    \end{tabular}
    }
    \caption{Results on LJSpeech for various end points for the first stage}
    \label{tab:results_timesteps_ljspeech}
\end{table}

\begin{table}[ht]
    \centering
    \resizebox{\columnwidth}{!}{
    \begin{tabular}{c|c|c|c}
         \textbf{Approach} & \textbf{PESQ} ($\uparrow$) & \textbf{STOI} ($\uparrow$) & \textbf{WARP-Q} ($\downarrow$) \\
         0 (GLA) & $3.824 \pm 0.246$ & $\mathbf{0.971} \pm 0.014$ & $\mathbf{1.082} \pm 0.083$ \\
         1 & $3.826 \pm 0.228$ & $0.936 \pm 0.052$ & $1.184 \pm 0.100$ \\
         2 & $\mathbf{3.830} \pm 0.216$ & $0.923 \pm 0.058$ & $1.325 \pm 0.100$ \\
         3 & $3.772 \pm 0.228$ & $0.917 \pm 0.057$ & $1.443 \pm 0.098$ \\
         4 & $3.651 \pm 0.244$ & $0.910 \pm 0.056$ & $1.476 \pm 0.099$ \\
         5 & $3.560 \pm 0.291$ & $0.909 \pm 0.056$ & $1.440 \pm 0.101$ \\
         6 (WaveGrad) & $3.453 \pm 0.325$ & $0.907 \pm 0.055$ & $1.439 \pm 0.100$ \\
    \end{tabular}
    }
    \caption{Results on VCTK for various end points for the first stage}
    \label{tab:results_timesteps_vctk}
\end{table}

\subsection{Impact of the end timestep of stage 1 per test file}\label{sec:impact_end_timestep_per_file}
We further investigate whether the optimal timestep varies across different test files. To this end, we compute the optimal timestep for each file and visualize the results using a histogram of the best timestep distribution for each metric.

For STOI and WARP-Q, we observe that GLA achieves better performance for most files on both VCTK and LJSpeech datasets. This superior average performance is also reflected at the individual file level.

\begin{figure}[ht]
    \centering
    \includegraphics[width=0.8\columnwidth]{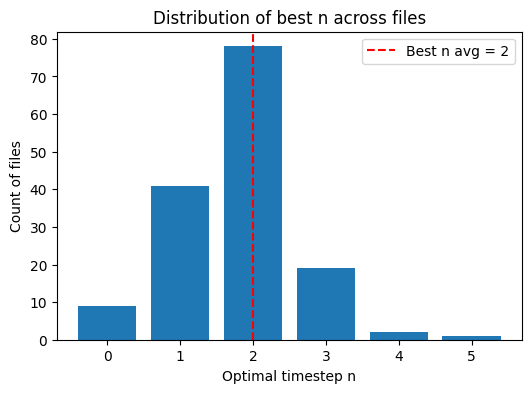}
    \caption{Histogram of the optimal timestep $n$ for PESQ across all files}
    \label{fig:histogram_ljspeech}
\end{figure}

For PESQ, the results for LJSpeech are shown in Figures~\ref{fig:histogram_ljspeech}, with similar trends observed for VCTK. Overall, the results indicate that the global optimal timestep is nearly optimal for about half of the files. However, the optimal end timestep can differ from one audio file to another. Notably, timestep 6 does not appear in the histogram, indicating that WaveGrad did not achieve the best PESQ for any file.

These findings suggest that guiding the diffusion process with the reconstructed signal $\tilde{\mbx}$ is consistently advantageous.

\section{Conclusion}\label{sec:conclusion}
In this paper, we propose GLA-Grad++, a zero-shot approach that leverages a reconstructed signal from the conditioning mel spectrogram to guide the diffusion process at a carefully chosen point within the denoising equation. Our experiments demonstrate that incorporating additional conditioning information improves the performance of diffusion vocoders, particularly for unseen speakers. We also show that the number of timesteps allocated to the correction stage has an effect on the results. Consequently, future work will focus on developing an algorithm to automatically select the optimal timestep for each individual file.

\clearpage
\bibliographystyle{IEEEbib}
\bibliography{refs}

\end{document}